\documentstyle[11pt,newpasp,twoside,epsf]{article}
\begin{document}
\title{Internal Motions in Globular Clusters}
\author{Ivan R. King and Jay Anderson}
\affil{Astronomy Department, University of California,\\ 
Berkeley, CA 94720-3411}

\def\hf{{\textstyle{1\over2}}}   
\def\subr #1{_{{\rm #1}}}
\def\avg#1{\langle #1 \rangle}
\def\hst{{\it HST\/}}
\font\boldital=cmbxti10 at 11pt
\def\eps@scaling{.95}
\def\epsscale#1{\gdef\eps@scaling{#1}}

\begin{abstract}
Observations of internal motions in globular clusters offer unique
insights into the dynamics of the clusters.  We have recently developed
methods of high-precision astrometry with \hst's WFPC2 camera, which
allow us to measure internal proper motions of individual stars.  These
new data open up many new avenues for study of the clusters.  Comparison
of the dispersion of proper motion with that of radial velocity offers
what is potentially the best method of measuring cluster distances, but
reliable results will require dynamical modeling of each cluster.
Proper motions are much better able to measure anisotropy of stellar
motions than are radial velocities.  In 47 Tucanae we have measured
thousands of proper motions near the center; their velocity distribution
is remarkably Gaussian.  In two outer fields we have begun to study
anisotropy, which appears in the high velocities but not in the lower
ones, contrary to the spheroidal velocity distributions that have
commonly been assumed.
\end{abstract}

\section{Introduction}

The dynamics of globular clusters is a venerable field, but new advances
continue to be made.  In this brief review we will deal with vistas that
have opened up as a result of work in which we are currently engaged.
Although we will deal with some theory, our discussion will be mainly
observational.

\section{{\boldital HST} astrometry}

We began trying to measure proper motions with the {\it Hubble Space
Telescope\/} (\hst) in order to separate cluster stars from field stars
in NGC 6397, where the cluster main sequence gets lost among the field
stars a magnitude and a half above our faint limit.  With WFPC2 images
spanning less than 3 years we found that we were able to effect an
excellent separation, and thereby detect the lowest part of the main
sequence (King et al.\ 1998).

\begin{figure}
\epsfxsize=8cm
\plotone{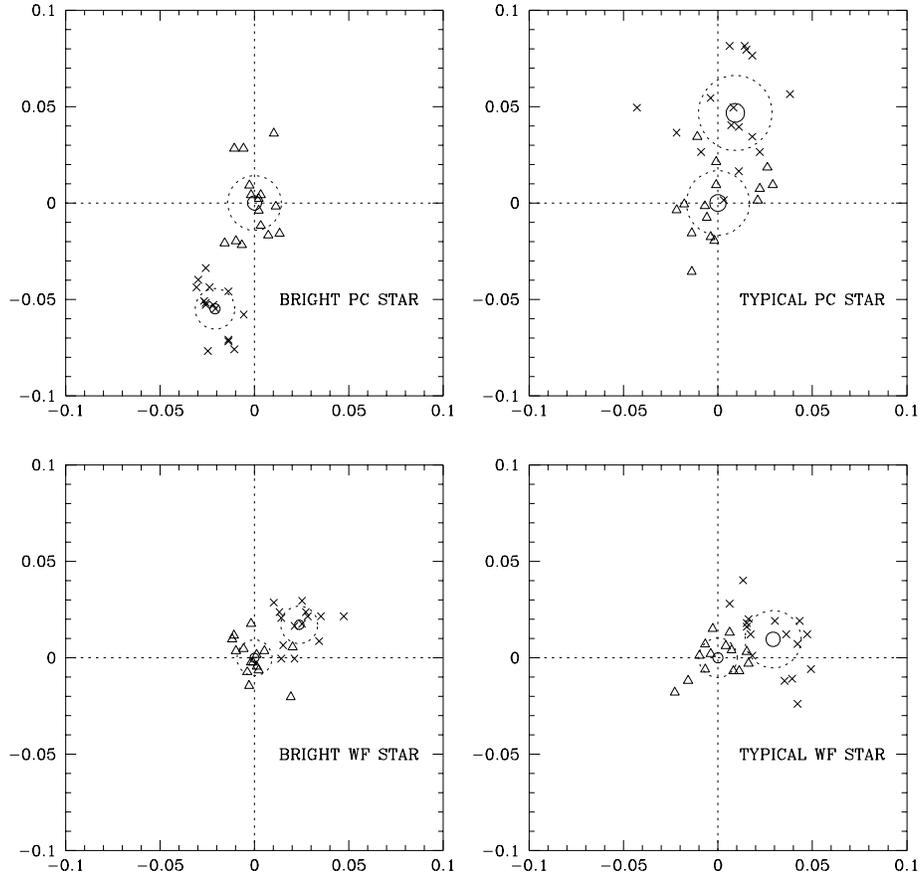}
\caption{Individual measurements and mean positions for stars in
47 Tuc.  Triangles are 1st epoch, crosses 2nd epoch.  Circles have radii equal
to the single-coordinate sigma, solid-line for the mean and dotted for the
scatter of individual measurements around the mean.  Unit is 1 pixel in each
camera.}
\label{displ}
\end{figure}

But the real turning point came when Georges Meylan asked us to measure
his WFPC2 images of the center of 47 Tucanae, taken two years apart.
His original aim was merely to detect which stars had the largest
internal proper motions in the cluster, but we quickly found that
our techniques could produce a motion for each individual star.  After
two more years of work on our methods, we have gained an improvement of
nearly an order of magnitude in our accuracy, by identifying and
removing the sources of systematic error that arise in the severely
undersampled WFPC2 camera.  The most important step, we have found, is
to derive an extremely accurate point-spread function (Anderson \& King
2000).

In 47 Tuc, where the dispersion of proper motions is about 0.6
milli-arc\-sec/yr in each coordinate, we are now in process of using
1995, 1997, and now 1999 images, to measure proper motions of individual
stars at about an $8\sigma$ level of accuracy or better.  At the time of
this conference we have derived proper motions only from the 1995 and
1997 images; they are at about the $4\sigma$ level.  Figure \ref{displ}
shows examples of our measurements and their scatter about the mean
positions that we derive from them.

The ability to measure proper motions of individual stars in globular
clusters opens up many new avenues of research, several of which we will
be discussing here.  We will examine the shape of the (isotropic)
velocity distribution at the cluster center, point out the value of
proper motions in measuring cluster distances, and, from another data
set, examine the anisotropy of the velocity distribution away from the
center.

\subsection{Velocity distribution in the cluster center}
\label{centvels}

At the center of the cluster there is no anisotropy in the distribution
of stellar velocities; as we shall indicate below, anisotropy manifests
itself only in the outer parts of a cluster.  What we can study quite
well here is the form of the velocity distribution.  We have motions for
enough stars to make a good comparison with the distribution that should
be expected.  We show such a comparison in Figure 2.  Most cluster
models have used a ``lowered Gaussian'', i.e., a Gaussian minus a
constant, where the constant is chosen so as to place the cutoff at the
proper escape velocity for the cluster center; we have made such a
choice here.  The only free parameter in the fit is the velocity
dispersion of the Gaussian.  Because even the 490 stars used here would
give rather noisy numbers if binned into a probability-distribution
histogram, we have chosen instead to exhibit a cumulative-distribution
step-function, which rises to the left because we have accumulated from
high velocities toward low.

The quantity that we have chosen to compare is the magnitude of the
proper motion.  The fit is strikingly good; the deviations near the
high-velocity end are probably due to small-number statistics.

\begin{figure}
\plottwo{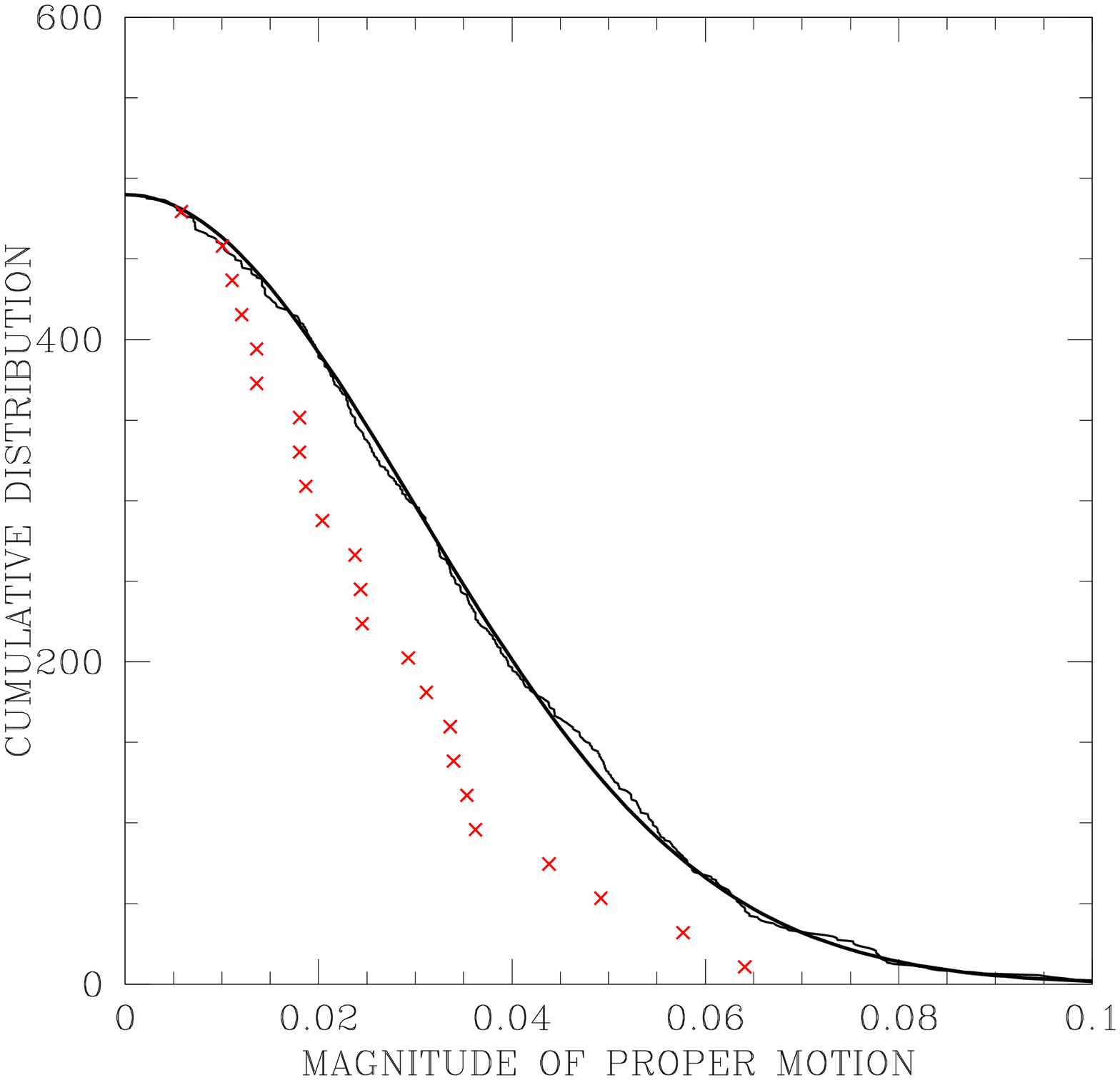}{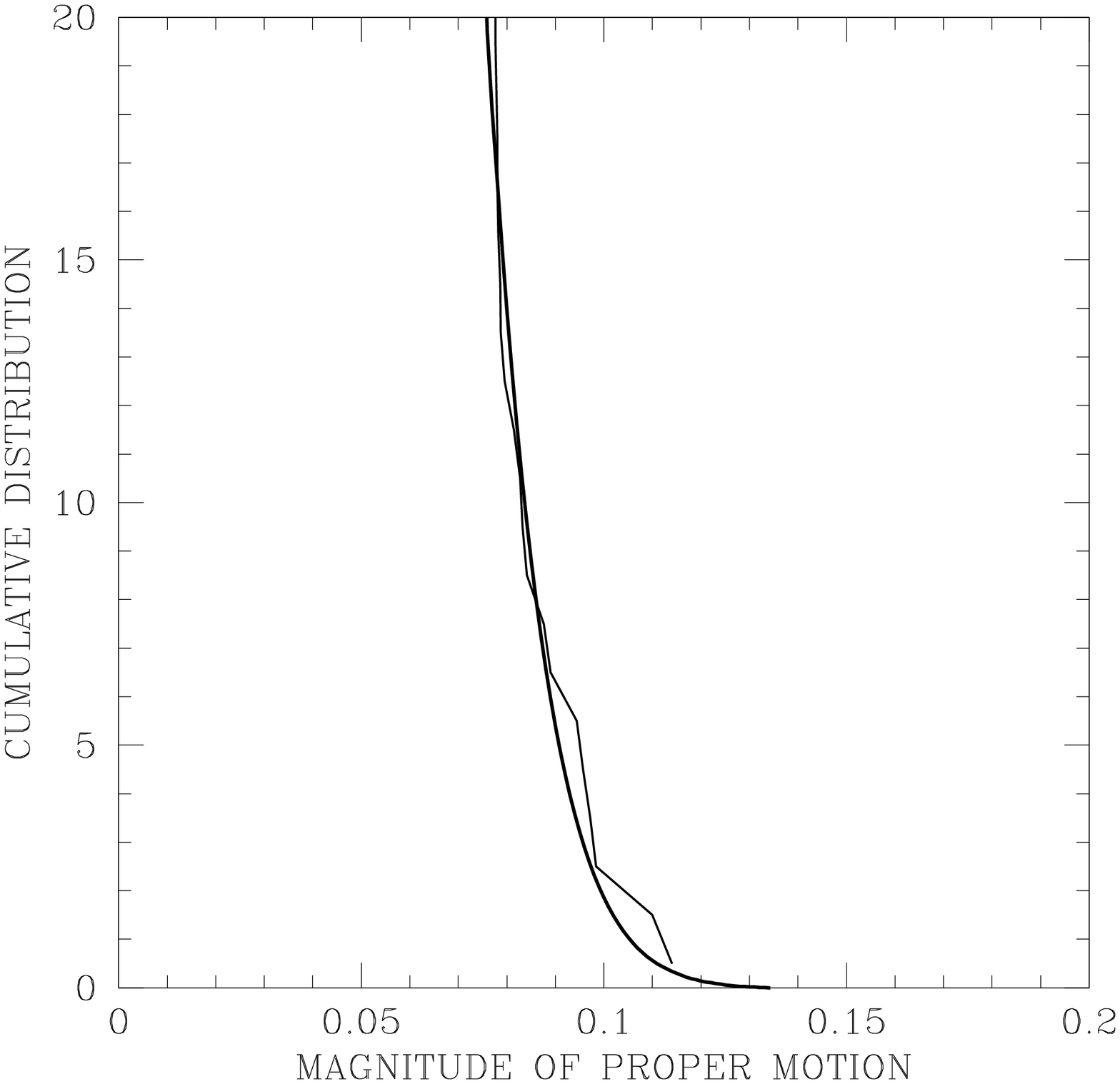}
\caption{Cumulative velocity distributions, accumulated from high
velocity toward low.  In each figure the solid line comes from a lowered
Gaussian that corresponds to the potential well at the center of 47 Tuc,
while the irregular line is accumulated from the magnitudes of the
proper motions of 490 RGB and AGB stars (down to $V$ magnitude 17.0).
The left-hand graph shows the whole distribution, while the right-hand
graph is a blowup of the region of the cutoff.  The crosses in the
left-hand graph show the distribution of the motions of blue stragglers,
which clearly have a lower velocity dispersion.}
\end{figure}
 
Interestingly, we have also been able to examine the ``cannonball
stars'' (Meylan, Dubath, \& Mayor 1991) whose existence was part of the
original motivation for measuring proper motions in 47 Tucanae.  They
turn out to have nearly zero proper motions, so that their radial
velocities are the whole of their motions, which do not appear to be
above the escape velocity at the center of the cluster.

Another distribution that we can examine is that of the blue stragglers.
They are known to be more concentrated to the cluster center than are
the red giants; they should therefore have a lower velocity dispersion.
Their distribution in Fig.\ 2 confirms this quite well.

\subsection{Cluster distances}
\label{dist}

Although it is outside the immediate focus of this conference, we should
mention that the most valuable product of such proper motions will be
fundamental distances of globular clusters, found by comparing the
dispersion of proper motions, an angular quantity, with that of radial
velocities, a linear quantity.  With 10,000--20,000 proper motions in a
cluster and at least 2500 radial velocities measured by our collaborators
(see, e.g., Gebhardt et al.\ 1997), the only completely unavoidable
error will be that due to statistical sample size, which is between 1
and 2\%.  Our aim, then, will be to avoid all other sources of error.
The errors of measurement are not a problem {\it per se\/}; what counts
is that their size be accurately known, so that they can be correctly
subtracted in quadrature in order to derive the true dispersions.

A serious concern will be the effect of orbital motion in binaries; it
must be corrected for statistically, and will be different for the
radial velocities and the proper motions, because of the different
distribution of the observations in time.  To make it worse, the binary
fraction in globular clusters is poorly known.

Another problem of consequence takes us back to dynamics:\ the radial
velocities of 47 Tuc stars show a quite prominent rotation of the
cluster (Meylan \& Mayor 1986).  We can in fact see rotation clearly in
the proper motions too; at the cluster center they show a slightly
greater dispersion in the equatorial than in the polar direction.  It is
obviously essential that we have a good dynamical model for the cluster.
(It should also go without saying that the model should have the correct
mass function for this cluster, a precept that has not always been
followed.  Fortunately our \hst\ observations of the cluster provide a
good mass function down to quite low masses.)  For the modeling itself
we intend to generalize the rotating models of Wilson (1975) to the
multi-mass case.

\section{The theory of anisotropy in globular clusters}
\label{theor_aniso}

\subsection{The origin of anisotropy}
\label{origin}
The most simple-minded dynamical models of clusters, such as the
original ``King models'' (King 1966), have isotropic velocity
distributions, but this is a serious oversimplification for real
clusters.  Whether a cluster is born by the collapse of a larger
configuration or by expansion of a smaller one, the initial motions are
largely radial; and conservation of angular momentum in each stellar
orbit will then constrain the motions to remain largely radial.  This
initial anisotropy should be removed gradually as relaxation moves the
velocity distribution toward isotropy, and this trend should spread
gradually outward from the dense center to the lower-density envelope.

Intuition suggests that the velocity distribution in a cluster should
have retained its anisotropy outside the radius where the local
relaxation time equals the cluster age (i.e., about a Hubble time), and
this is approximately what is observed.

\subsection{The nature of anisotropy}
\label{nature}
In a spherically symmetric globular cluster, the three phase-space
variables on which the velocity distribution can depend are the radial
distance $r$, the radial component of velocity $v\subr{r}$, and the
tangential component of velocity $v\subr{t}$.  It has been customary to
represent anisotropy in a globular cluster by writing the velocity
distribution as
\begin{equation}
f(r,v\subr{r},v\subr{t}) = F(E + cJ^2),
\end{equation}
where 
\begin{equation}
E=\hf\left(v\subr{r}^2 + v\subr{t}^2\right) + U(r)
\end{equation}
is the energy per unit mass ($U$ being, of course, the gravitational
potential function), and
\begin{equation}
J=r v\subr{t}
\end{equation}
is the magnitude of the angular momentum per unit mass. The reasons for
this representation are
that (1) Jeans' theorem says that in a time-independent state (which is
a very good approximation in a slowly evolving system) the velocity
distribution must be a function of these two integrals of the orbital
motion of a star, and (2) this simple form is easily shown to lead to a
velocity spheroid whose radial elongation increases with increasing $r$.

It is this kind of velocity distribution that Gunn and Griffin (1979)
and Meylan (1987, 1988) used in their studies of anisotropy, and a
modified form of this distribution that Lupton, Gunn, \& Griffin (1987)
used in their treatment of anisotropy in a rotating cluster.
We will show below, however, that observed anisotropic velocity
distributions unfortunately do not have this form.

\subsection{Dependence of anisotropy on stellar mass}
\label{massdep}

Nothing is known about how anisotropy differs from one stellar mass to
another.  We are not aware of any theoretical prediction, nor is there
observational information, since existing studies, through radial
velocities, have been confined to stars above the main-sequence turnoff,
all of which have practically the same mass.  We expect to derive such
information from observations that will be described below, but for the
time being one can only make a rough theoretical estimate.

Anisotropy is removed by relaxation, whose rate goes as $v^{-3}$, where
$v$ is a characteristic velocity for the stars in question.  Since in
the condition of equipartition that seems to exist in nearly all
globular clusters $v\propto 1/m$, the time to remove anisotropy would at
first appear to go as $m^{-3/2}$.  But the dependence on mass should be
even stronger, since the stars of lower mass inhabit regions of lower
density, on account of their higher velocities, and this makes their
relaxation even slower.  A quantitative study, however, promises to be
rather complicated, and we will not pursue it here.

\section{Methods of observing anisotropy}
\label{obs_aniso}

Until now, anisotropy of stellar motions in globular clusters has been
studied only indirectly, by comparing the radial drop-off that is
observed in the dispersion of radial velocities with the drop-off that
is predicted by theoretical models with differing amounts of anisotropy.
Not only is this method indirect; it is also too vulnerable to
inadequacies in the dynamical modeling.  By contrast, proper-motion
measurements measure anisotropy directly, and are nearly independent of
any modeling.

The analysis of the radial-velocity observations depends on our
different geometrical perspective at the center and at the edge of the
cluster.  The velocities that we observe along the line of sight at the
cluster center are clearly values of $v\subr{r}$ only, whereas at points
away from the center the observed dispersion has an increasing admixture
of $v\subr{t}$, so that the drop-off of ${\avg{v^2}}\subr{los}$ gives us
information about the anisotropy.  Unfortunately that information is
only indirect.  The problem is that there is a natural drop-off of
${\avg{v^2}}\subr{los}$ even in a cluster whose velocity distribution is
isotropic; because the velocity distribution has a cutoff at the escape
velocity, the decrease of the latter with increasing radius causes
${\avg{v^2}}\subr{los}$ to fall.  Thus the observed quantity that is
needed becomes in effect a higher-order one---the difference between the
actual fall-off of ${\avg{v^2}}\subr{los}$ and the fall-off that would
occur in a cluster model that has isotropic velocities.  This weakness
gives radial velocities only a poor grip on anisotropy.  One should note
further that such a measurement of anisotropy is very sensitive to the
cluster model adopted.  The results will suffer badly from any errors
either in the assumed mass function or in the functional form that is
used for the representation of anisotropy in the velocity distribution.
In the latter respect, we will show below that the functional form that
has commonly been used is contradicted by our new observations.

The proper motions play a quite different role.  At a point away from
the cluster center one can directly compare the dispersion in the radial
direction with that in the transverse direction.  There are, to be sure,
projection problems along the line of sight (as there are equally in the
interpretation of the radial velocities), but the proper motions give us
our only really direct measure of anisotropy.

A caution to note, however, is that rotation of the cluster will distort
the apparent anisotropy.  Along a given line of sight, components of the
rotational motion will be included in the part of the proper motions
that is parallel to the cluster equator.  (This is similar to the small 
rotation-induced anisotropy that we have already noted at the cluster
center.)  Interpretation of the proper motions thus has a small
dependence on modeling of the cluster, but far less than in the case of
the radial velocities.

\section{Our new observations:\ anisotropy in 47 Tuc}
\label{obsaniso}

In addition to our work on Meylan's images at the center of 47 Tucanae,
we have a program of our own, in which we have images of two fields on
opposite sides of the center of 47 Tuc, each at about 10 core radii from
the center.  Meylan's radial-velocity study (1988) suggests that
anisotropy becomes strong at a radius somewhere between there and 40
core radii, so our fields should show a modest amount of anisotropy.
Our time baselines are 4--5 years.  As already indicated, we are able to
examine anisotropy directly, by comparing proper-motion dispersions in
the radial and transverse directions.  We have so far made a preliminary
reduction of one of these fields.  Although our measurements are still
only preliminary, their trend is clear.

The results are surprising.  We do see anisotropy, but {\it only among
the stars of highest velocity\/}.  In other words, in our
two-dimensional distribution of proper motions the lines of constant
density are not ellipses of constant axial ratio.  The contours near the
center are circles, while only the largest contours show an elongation
in the radial direction.  These observations contradict the picture of
similar-shaped ellipses that would follow from the sort of velocity
distribution that is specified by Eq.\ (1).

This is a result that we had not foreseen, and as far as we know no one
else had either; but it should not have been a surprise, because in
clear hindsight it is exactly what we should have expected.  To
understand it we must keep in mind that anisotropy is removed only by
relaxation, whose rate is proportional to the star density in the regions
through which a star passes.  In a region like the one that we have
observed, the stars of low velocity have orbits that bring them into the
denser inner regions, where they have seen enough relaxation to
isotropize their orbits, whereas those of higher velocity have orbits in
which they have spent most of their lives in the low-density outer
regions, and with less exposure to the relaxing tendency they have
retained some of their anisotropy.

Given the fact that anisotropic velocity distributions are not as we
thought they were, how is the theory of cluster dynamics to react?
Obviously we need to construct cluster models based on velocity
distributions in which the roles of $E$ and $J$ differ from the
simplistic ones in Eq.\ (1).  Unfortunately it is not yet clear what
sort of mathematical function will fill the bill.  An important
challenge to the building of realistic dynamical models of star clusters
is to find a suitable function.  Presumably, for the sake of providing a
finite escape velocity it can then be cast in a form analogous to the
\begin{equation}
f(E,J^2) = e^{-2j^2(E+aJ^2)} - e^{-2j^2 aJ^2},
\end{equation}
which is the form used in the now-outmoded models.

We look forward to a better understanding of anisotropy after we have
fully reduced the images of these two fields in 47 Tuc, and especially
after measuring an outer field in $\omega$ Centauri, where from Meylan's
(1987) results we expect the anisotropy to be much greater.

The interpretation of anisotropy observations will also involve modeling
of the cluster.  In this case we will need models that include both
anisotropy and rotation.  Fortunately Lupton \& Gunn (1987) have shown
how to build such models.

\section{Summary}

Observations of internal motions in globular clusters are obviously 
essential for understanding the dynamics of these systems.  Radial 
velocities have contributed some information, but the new ability to 
measure internal proper motions of individual stars, using the {\it 
Hubble Space Telescope}, opens new vistas for observational studies of 
dynamics.

Proper motions of large numbers of stars show that the velocity 
distribution is quite close to Gaussian, as dynamical models have 
supposed.

Proper motions are much more suitable than radial velocities for the
study of anisotropy of velocity dispersions in the outer parts of
clusters.  Here preliminary results indicate that the anisotropy does
not take the form that is customarily assumed in dynamical models;
hindsight easily shows why this should be so.  New mathematical forms
for the velocity distribution need to be contrived.

The most important application of internal proper motions in globular 
clusters will be the derivation of fundamental distances, from a 
comparison of the dispersion of the proper motions with that of the 
radial velocities.

Finally, we note that in a later paper in this volume Ken Freeman 
reports results of a unique ground-based study that analyzes internal 
proper motions in $\omega$ Centauri.

\acknowledgments
The research supported here was supported by grant GO-7503 from the
Space Telescope Science Institute.


\begin{references}

\reference Anderson, J., \& King, I.\ R. 2000, \pasp, in press
               
\reference Gebhardt, K., Pryor, C., Williams, T. B., Hesser, J.\ E., \& 
Stetson, P.\ B. 1997, \aj, 113, 1026

\reference Gunn, J.\ E., \& Griffin, R.\ F. 1979, \aj, 84, 752

\reference King, I.\ R. 1966, \aj, 71, 66

\reference King, I.\ R., Anderson, J., Cool, A.\ M., \& Piotto, G. 1998,
\apjl, 492, L37 


\reference Lupton, R., \& Gunn, J. 1987, \aj, 93, 1106

\reference Lupton, R., Gunn, J., \& Griffin, R. 1987, \aj, 93, 1114

\reference Meylan, G. 1987, \aap, 184, 144

\reference Meylan, G. 1988, \aap, 191, 215

\reference Meylan, G., \& Mayor, M. 1986, \aap, 166, 122

\reference Meylan, G., Dubath, P., \& Mayor, M. 1991, \apj, 383, 587

\reference Wilson, C.\ P. 1975, \aj, 80, 175

\end{references}
\end{document}